\documentstyle[twocolumn,graphicx,prb,aps]{revtex}
\input epsf
\tighten
\draft
\preprint{}
\begin{document}

\twocolumn[\hsize\textwidth\columnwidth\hsize \csname
@twocolumnfalse\endcsname 
\title{Isotope effect in the presence of a pseudogap}

\author{T.~Dahm \cite{address}}

\address{\sl Max-Planck-Institute
         for Physics of Complex Systems, 
         N\"othnitzer Str.~38, 
         D-01187 Dresden, Germany}

\address{\sl Institut f\"ur Theoretische Physik, 
         Universit\"at T\"ubingen, 
         Auf der Morgenstelle 14, D-72076 T\"ubingen, 
         Germany}

%\date{Draft made \today}
\maketitle

\begin{abstract}
  We investigate the question whether the unusual doping dependence
of the isotope exponent observed in underdoped high-$T_c$ superconductors
might be related to another unusual phenomenon observed in these
systems: the pseudogap phenomenon. Within different approximations
we study the influence of a phenomenological pseudogap on the
isotope exponent and find that it generally strongly increases the
isotope exponent, in qualitative agreement with experiments on
underdoped high-$T_c$ compounds. This result is stable against
strong-coupling self-energy corrections and also holds for
recently proposed spinfluctuation exchange models, if a weak
additional electron-phonon coupling is considered.
\end{abstract}

]

\section{Introduction}
\label{secI}

The isotope effect in high-$T_c$ cuprate superconductors is
unconventional in different respects. Optimally doped samples
show a very small isotope exponent $\alpha$ of the order of
0.05 or even smaller, in contrast to the conventional 
Bardeen-Cooper-Schrieffer (BCS) value
of 0.5, which one expects for a conventional phonon induced
pairing interaction. \cite{Franck} This unusually small value in
connection with the high value of $T_c$ lead to early 
suggestions that the pairing interaction in high-$T_c$ cuprates
might be predominantly electronic in origin with a possible small
phononic contribution. \cite{Marsiglio} This scenario, however,
is difficult to reconcile with the fact that the isotope
coefficient also shows an unusually strong doping dependence,
reaching values of 0.5, in some cases even higher, in the
underdoped, $T_c$ reduced, compounds. \cite{Franck,Franck2,DMTiso}

Many different models have been advanced in order to try to
understand this unusual doping dependence in connection with
the small isotope exponent at optimal doping, e.g. influence
of van Hove singularities \cite{Labbe,Tsuei,Radtke}, anharmonic phonons
\cite{Pao,Pietronero}, electron-phonon coupling in the presence of
strong antiferromagnetic correlations \cite{Nazarenko,Greco},
pair breaking effects \cite{Carbotte},
magnetic impurities or Jahn-Teller nonadiabaticity \cite{Bill,Bill2},
but no consensus has been reached so far. \cite{refs}

In recent years it became apparent that the physics of underdoped 
high-$T_c$ superconductors is governed by the pseudogap phenomenon. A
behavior which is reminiscent of the presence of a pseudogap,
growing upon successive underdoping, has been observed consistently
in a large number of different experiments, e.g. nuclear magnetic
resonance (NMR) Knight-shift and
relaxation rate experiments, specific heat, angular-resolved
photoemission spectroscopy (ARPES), tunneling, $c$-axis and $ab$-plane
dynamical conductivity \cite{Timusk,Williams}.
Currently there is no consensus about the origin of this pseudogap
and many different proposals exist. 
\cite{Emery,Zhang,Pines,Lee,Alexandrov,Maki,Zachar,DMT}
Williams et al. \cite{Williams} have shown that a phenomenological model 
for a pseudogap having $d$-wave symmetry can account well for thermodynamic
quantities in the underdoped cuprates.

In the present manuscript we want to study whether the influence
of the pseudogap might give a, perhaps more natural, explanation
for the unusual isotope effect in underdoped high-$T_c$ cuprates.
As has been shown by Carbotte et al. \cite{Lie,Schachinger} an
energy dependence of the electronic density of states (DOS) varying
on the pairing energy scale can modify the isotope effect, and therefore
we ask whether there might be a link between the pseudogap
phenomenon and the isotope effect. Since there exists no widely
accepted theory for the pseudogap at present, here we will follow
the idea of Williams et al and treat the pseudogap on a
phenomenological basis, introducing it into the single particle
excitation spectrum. Such a procedure is reasonable, if the
pseudogap itself does not show an isotope effect, as suggested
by recent NMR experiments \cite{Williams2,Raffa}.

In the following we will study the influence of such a pseudogap
on the isotope effect within different models. We shall start with 
the weak-coupling (BCS)
approximation where we consider $s$- and $d$-wave symmetry of the
superconducting order parameter and of the pseudogap. In order to
see whether these results are stable for more realistic cases, we 
will study two recently proposed models based on a spinfluctuation exchange
pairing interaction, e.g. the nearly antiferromagnetic Fermi liquid (NAFL)
model due to Monthoux and Pines \cite{Monthoux} and the self-consistent
fluctuation-exchange (FLEX) approximation for the two dimensional
Hubbard model. \cite{Bickers}

\section{Weak coupling approximation}
\label{secweak}

The linearized gap equation in the weak-coupling limit
for an anisotropic pairing interaction $V(\vec{k},\vec{k}^\prime)$
reads:

\begin{equation}
 \Delta ( \vec{k} ) = \frac{1}{N} \sum_{k^\prime} 
 V ( \vec{k},\vec{k}^\prime )
 \frac{\tanh \left( \epsilon_{k^\prime} / 2 T_c \right)}
  {2 \epsilon_{k^\prime}} \Delta ( \vec{k}^\prime )
\label{eq1}
\end{equation}
Here, $\epsilon_k$ is the band dispersion and $\Delta ( \vec{k} )$
the superconducting gap function.
We want to assume that the pairing interaction consists of two
parts: a phononic part $V_p(\vec{k},\vec{k}^\prime)$ and an
electronic part $V_e(\vec{k},\vec{k}^\prime)$, such that
$V(\vec{k},\vec{k}^\prime)=V_p(\vec{k},\vec{k}^\prime)
+ V_e(\vec{k},\vec{k}^\prime)$. The dominant contribution
shall be $V_e$. Then, in weak-coupling approximation we have
\begin{equation}
 V_e ( \vec{k},\vec{k}^\prime ) = \left\{
  \begin{array}{c@{\quad}l} V_{e0} \psi_\eta ( \vec{k} )
  \psi_\eta ( \vec{k}^\prime ) & {\rm if} \quad |\epsilon_k|,
  |\epsilon_{k^\prime}| \le \omega_e \\ 0 & {\rm else}
  \end{array} \right. ,
\label{eq2}
\end{equation}
where $\omega_e$ is the characteristic energy scale of the electronic
part and is assumed to be independent of isotopic mass. 
$\psi_\eta ( \vec{k} )$ is the basis function for the pairing symmetry
considered. For $s$-wave pairing $\psi_s ( \vec{k} )=1$, for
$d_{x^2-y^2}$-wave pairing $\psi_{d_{x^2-y^2}} ( \vec{k} )=\cos 2\Theta_k /
\sqrt{2}$ and for $d_{xy}$-wave pairing $\psi_{d_{xy}} ( \vec{k} )=
\sin 2\Theta_k / \sqrt{2}$, where $\Theta_k=\arctan (k_y/k_x)$ is
the angular direction of the momentum $\vec{k}$.

The phononic part may consist of different contributions having
different symmetries. However, since we assumed that the electronic part
is dominating with a symmetry specified by $\psi_\eta$, only the
$\psi_\eta$-component of $V_p$, having the same symmetry, will affect
$T_c$. Therefore we can assume without loss of generality
\begin{equation}
 V_p ( \vec{k},\vec{k}^\prime ) = \left\{
  \begin{array}{c@{\quad}l} V_{p0} \psi_\eta ( \vec{k} )
  \psi_\eta ( \vec{k}^\prime ) & {\rm if} \quad |\epsilon_k|,
  |\epsilon_{k^\prime}| \le \omega_p \\ 0 & {\rm else}
  \end{array} \right. ,
\label{eq3}
\end{equation}
where $\omega_p$ is the characteristic phonon energy. In harmonic
approximation, which we will adopt here, $\omega_p$ varies with
the isotopic mass $M$ like $1/\sqrt{M}$, while $\omega_e$ is
assumed to be independent of $M$.

For such an interaction the gap function can be separated
into two parts: $\Delta ( \vec{k} ) = \Delta_e ( \vec{k} ) 
+ \Delta_p ( \vec{k} ) $, with
\begin{equation}
 \Delta_{e,p} ( \vec{k} ) = \left\{
  \begin{array}{c@{\quad}l} \Delta_{e0,p0} \psi_\eta ( \vec{k} )
  & {\rm if} \quad |\epsilon_k| \le \omega_{e,p} \\ 0 & {\rm else}
  \end{array} \right. .
\label{eq4}
\end{equation}
With this ansatz Eq. (\ref{eq1}) becomes a $2 \times 2$ matrix
equation for the two order parameter components $\Delta_{e0}$
and $\Delta_{p0}$. Assuming a cylindrical Fermi surface with
a constant density of states Eq. (\ref{eq1}) can be written
in the form
\begin{equation}
 \left( \begin{array}{c} \Delta_{e0} \\  \Delta_{p0} \end{array}  \right)
  = \left(
  \begin{array}{cc} V_{e0} L(\omega_e) & V_{e0} L(\omega_e) \\
  V_{p0} L(\omega_e) & V_{p0} L(\omega_p) \end{array} \right)
  \left( \begin{array}{c} \Delta_{e0} \\  \Delta_{p0} \end{array}  \right) ,
\label{eq5}
\end{equation}
where we defined the function $L(\omega)$
\begin{equation}
 L(\omega) = 
 N(0) \int_0^\omega d\epsilon \frac{\tanh \left( \frac{\epsilon} {2 T} \right)}
  {\epsilon}  \simeq N(0) \ln \left( \frac{1.13 \omega}{T} \right)
\label{eq6}
\end{equation}
The last expression holds in the weak coupling limit $\omega \gg T$.
$N(0)$ denotes the density of states at the Fermi level.
In deriving Eq. (\ref{eq5}) we assumed $\omega_e \le \omega_p$,
as is usually the case for spinfluctuation exchange models (see
the following sections).
However, the final result Eq. (\ref{eq13}) does not depend
on this choice.
Letting $L_p=L(\omega_p)$ and $L_e=L(\omega_e)$
the leading eigenvalue of the matrix in Eq. (\ref{eq5}) is
\begin{eqnarray}
 \lambda (\omega_e,\omega_p,T) & = & \frac{V_{e0} L_e + V_{p0} L_p}{2} 
 + \nonumber \\ & & + \frac{1}{2} \sqrt{(V_{e0} L_e - V_{p0} L_p)^2 + 
                4 V_{e0} V_{p0} L_e^2 }
\label{eq7}
\end{eqnarray}
and $T_c$ is determined from the implicit equation
\begin{equation}
 \lambda (\omega_e,\omega_p,T_c) = 1 .
\label{eq8}
\end{equation}
From this the isotope exponent $\alpha_0$ can be calculated:
\begin{equation}
 \alpha_0= \frac{1}{2} \frac{d \ln T_c}{d \ln \omega_p}
 = - \frac{1}{2} \frac{\omega_p}{T_c} \frac{\frac{\partial \lambda}{\partial
  L_p} \frac{\partial L_p}{\partial \omega_p}}
  {\frac{\partial \lambda}{\partial L_p} \frac{\partial L_p}{\partial T_c}
  + \frac{\partial \lambda}{\partial L_e} \frac{\partial L_e}{\partial T_c} }
\label{eq9}
\end{equation}
In the weak coupling limit $\omega_p,\omega_e \gg T_c$ this gives
\begin{equation}
 \alpha_0 = \frac{1}{2} \frac{V_{p0} (1-V_{e0} L_e)}{V_{p0} (1+V_{e0} L_e) +
V_{e0} (1 - V_{p0} L_p)} .
\label{eq10}
\end{equation}
Note, that for a purely electronic interaction $V_{p0}=0$ this expression
yields $\alpha_0=0$ and for a purely phononic interaction $V_{e0}=0$ it gives
$\alpha_0=0.5$, as one should expect. For a mixed interaction $\alpha_0$ will
generally lie between 0 and 0.5. In fact, one can easily show that for given values
of $\omega_p$ and $\omega_e$ one can always choose $V_{p0}$ and $V_{e0}$ in such a
way that a given value of $T_c$ and $\alpha_0 \in [0,0.5]$ is reached. \cite{Dahmunp}

\begin{figure}[t]
  \begin{center}
    \includegraphics[width=0.65\columnwidth,angle=270]{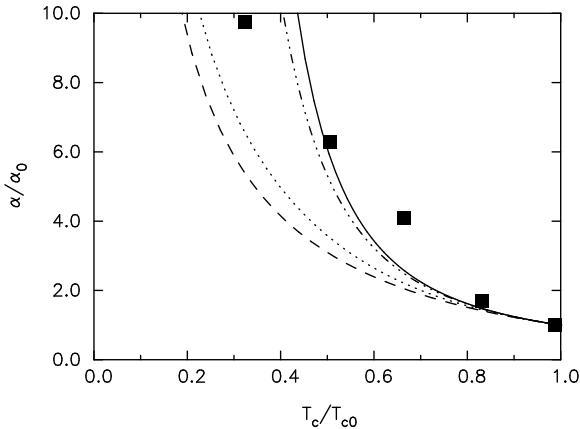}
    \vspace{.2cm}
    \caption{Weak-coupling result Eq. (\ref{eq13}) for the isotope exponent
     $\alpha/\alpha_0$ as a function of $T_c/T_{c0}$ in the presence of a
     pseudogap. $\alpha_0$ and $T_{c0}$ denote the values in the absence of the
     pseudogap. The solid line shows the result for an $s$-wave pseudogap. For
     a $d_{x^2-y^2}$-wave pseudogap the results for an $s$-wave pairing
     symmetry (dotted line), a $d_{x^2-y^2}$-wave pairing symmetry
     (dashed-dotted line), and a $d_{xy}$-wave pairing symmetry (dashed line)
     are shown. The solid squares are experimental results on Pr-doped YBCO
     from Ref. \protect\onlinecite{Franck}.
     \label{Fig1} }
  \end{center}
\end{figure}

Now we wish to consider the influence of a pseudogap. In the presence of
a pseudogap we have to modify the single particle excitation spectrum.
Following Williams et al, we replace in Eq. (\ref{eq1})
\begin{equation}
  \epsilon_k \quad \Longrightarrow \quad \sqrt{\epsilon_k^2 + E_g^2(\vec{k})} ,
\label{eq11}
\end{equation}
where $ E_g(\vec{k})$ is the pseudogap and will be chosen to be either
$ E_{g,s}(\vec{k})=E_{g0} = {\rm const}$ for an $s$-wave pseudogap or
$ E_{g,d}(\vec{k})=E_{g0} \cos 2 \Theta_k$ for a $d$-wave type
pseudogap. Note, that this symmetry of the pseudogap not necessarily has to
be identical with the pairing symmetry and we will allow them to be 
independent in this section. However, the study in Ref. \onlinecite{Williams}
suggests that both symmetries are of $d$-wave type in underdoped high-$T_c$
compounds and we will focus on this case in the following sections.
With the replacement Eq. (\ref{eq11}) the function $L$ becomes
\begin{equation}
 L(\omega) = \frac{N(0)}{2\pi} \int\limits_0^{2\pi} d\Theta \psi_\eta^2 ( \Theta ) 
 \int\limits_0^\omega d\epsilon \frac{\tanh \left( \frac{\sqrt{\epsilon^2 +
E_g^2(\Theta)}} {2 T} \right)}
  {\sqrt{\epsilon^2 + E_g^2(\Theta)}} .
\label{eq12}
\end{equation}
Eqs. (\ref{eq7}) and (\ref{eq9}) still remain valid, if one uses this
expression for $L(\omega)$. In the weak-coupling limit 
$\omega_p,\omega_e \gg T_c,E_g$ we then find for the isotope exponent
\begin{equation}
 \alpha = \alpha_0 \left( \frac{1}{4\pi T_c} \int\limits_0^{2\pi} 
  d \Theta
 \int\limits_0^\infty d\epsilon \frac{\psi_\eta^2 ( \Theta ) }
  {\cosh^2 \frac{\sqrt{\epsilon^2+E_g^2(\Theta)}}{2 T_c}}
  \right)^{-1}
\label{eq13}
\end{equation}
where $\alpha_0$ is the isotope exponent Eq. (\ref{eq10}) in the absence
of a pseudogap. Eq. (\ref{eq13}) shows that $\alpha/\alpha_0$ only depends
on $E_{g0}/T_c$, the pairing symmetry $\psi_\eta ( \Theta )$ and
the symmetry of the pseudogap. Since $T_c$ is a function of $E_{g0}$,
determined from Eq. (\ref{eq8}), for a given symmetry of both the
pseudogap and the pairing state $\alpha/\alpha_0$ is a {\it universal}
function of $T_c/T_{c0}$. Here, $T_{c0}=T_c(E_g = 0)$. In Fig.
\ref{Fig1} we show $\alpha/\alpha_0$ as a function of $T_c/T_{c0}$
for different symmetries. The solid line shows the isotope exponent
for an $s$-wave pseudogap. This result is independent of the pairing
symmetry, as can be seen by performing the angular integration in Eq. 
(\ref{eq13}).
For an anisotropic pseudogap having $d_{x^2-y^2}$-wave symmetry,
however, the pairing symmetry does affect the result. The dotted line 
shows the result
for an $s$-wave superconductor with a $d_{x^2-y^2}$-wave pseudogap,
while the dashed-dotted line shows the result for a $d_{x^2-y^2}$-wave
superconductor with a $d_{x^2-y^2}$-wave pseudogap. The weakest $T_c/T_{c0}$
dependence is found for a $d_{xy}$ superconductor with a 
$d_{x^2-y^2}$-wave pseudogap (dashed line). In all cases one can see
from Eq. (\ref{eq13}) that $\alpha/\alpha_0$ diverges for $T_c \rightarrow 0$.
Thus, in principle arbitrarily high values of $\alpha$ can be reached.
As an illustration experimental results on Pr-doped YBCO are
shown in this figure as solid squares. \cite{Franck}
Here, it should be noted that experimental results on different
compounds can differ somewhat and also vary with the dopant used 
(see Ref. \onlinecite{Franck}).
Certainly these differences need further explanation (e.g. see the
review in Ref. \onlinecite{Bill2}) and cannot
be understood solely due to the influence of the pseudogap. 
Here, we only want to focus on the influence of a pseudogap alone and
investigate the general tendency and order 
of magnitude of the effect, which is similar in many compounds.

As an important conclusion we can draw from these weak-coupling results that 
a pseudogap in
general leads to an increase of the isotope exponent $\alpha$ over its
value $\alpha_0$ in the absence of a pseudogap. The quantitative size of this
effect depends on the symmetries of the pseudogap and the pairing state.
However, the qualitative behavior is very similar in all cases. The size of
$\alpha_0$ can become small, if a strong electronic coupling constant
$V_{e0}$ and a small phononic coupling $V_{p0}$ is considered.

\section{Strong coupling effects: NAFL model}
\label{secnafl}

Having seen that a pseudogap can lead to an increase of the isotope
exponent in a weak-coupling superconductor, one might wonder whether
this effect will survive in more realistic models for superconductivity.
In order to see, how strong-coupling effects affect the results, we want
to consider a recently proposed spinfluctuation exchange model,
the nearly antiferromagnetic Fermi liquid (NAFL) model due to Monthoux
and Pines \cite{Monthoux}. Within this model the pairing interaction is
provided by exchange of antiferromagnetic spinfluctuations and the
pairing symmetry is $d_{x^2-y^2}$. The (frequency
dependent) pairing interaction is given by
\begin{equation}
  V(\vec{q},i\nu_m) = g^2 \chi(\vec{q},i\nu_m) \quad ,
\label{eq14}
\end{equation}
where $g$ is a coupling constant, $\nu_m$ the Bose-Matsubara frequencies,
and the spinsusceptibility $\chi$ is given by
\begin{equation}
  \chi(\vec{q},i\nu_m) = \frac{\chi_Q}{1+\xi^2 (\vec{q}-\vec{Q})^2
  + \nu_m/\omega_{s} } 
\label{eq15}
\end{equation}
Here, $\vec{Q}=(\pi,\pi)$ is the antiferromagnetic wavevector, $\xi$
is the magnetic correlation length and $\omega_s$ the characteristic
spinfluctuation frequency.
Using this interaction the Migdal-Eliashberg equations for
strong-coup\-ling superconductors are solved selfconsistently.
Here one has to solve for the self-energy $\Sigma$
\begin{equation}
\Sigma(\vec{k},i\omega_n) = \frac{1}{N} \sum_{k^\prime,n^\prime}
V(\vec{k}-\vec{k}^\prime,i\omega_n-i\omega_{n^\prime})
G(\vec{k},i\omega_n) 
\label{eq16}
\end{equation}
along with Dyson's equation for the Green's function $G$
\begin{equation}
G(\vec{k},i\omega_n)= \frac{1}{i\omega_n-\epsilon_k-
\Sigma(\vec{k},i\omega_n)} 
\label{eq17}
\end{equation}
selfconsistently. Using this solution, $T_c$ is determined
from the linearized gap-equation
\begin{eqnarray}
\phi(\vec{k},i\omega_n) & = & -\frac{1}{N} \sum_{k^\prime,n^\prime}
V(\vec{k}-\vec{k}^\prime,i\omega_n-i\omega_{n^\prime})
\nonumber \\ & & \times |G(\vec{k},i\omega_n)|^2 \phi(\vec{k},i\omega_n) \quad ,
\label{eq18}
\end{eqnarray}
where $\phi$ is the gap-function. For the bandstructure $\epsilon_k$
a tight-binding band with next nearest neighbor hopping has been
used in Ref. \onlinecite{Monthoux} and we will adopt that here.

In order to have a small nonzero
isotope exponent at optimal doping we consider coupling to an
additional phonon mode, the 'buckling' mode studied in Refs.
\onlinecite{Nazarenko,Song}, and \onlinecite{Bulut}. This mode 
provides an attraction in the $d_{x^2-y^2}$-wave channel and its
pairing interaction reads
\begin{equation}
V_p(\vec{q},i\nu_m) = V_{p0} ( \cos^2 \frac{q_x}{2} + \cos^2 \frac{q_y}{2} )
\frac{\omega_p^2}{\nu_m^2 + \omega_p^2} 
\label{eq19}
\end{equation}

We do not expect the main results to depend strongly
on the details of the electron-phonon spectrum, as long as its coupling
strength is small compared with the spinfluctuation interaction.
It is important, however, that the electron-phonon interaction
has an attractive component in the $d_{x^2-y^2}$-wave channel, as 
has been discussed above Eq. (\ref{eq3}).
For the calculations we choose the parameters given by
Monthoux and Pines: $\xi=2.3 a$, $\chi_Q=44 \; {\rm states/eV}$,
hopping matrix element $t=250 \; {\rm meV}$, where $a$ is the
lattice constant. \cite{Monthoux} For the spinfluctuation frequency 
$\omega_s$ we choose three different values 0.03$t$, 0.06$t$, and
0.2$t$, in order to study the cross-over from weak-coupling to
strong-coupling behavior. For the characteristic phonon
frequency we choose a typical value of $\omega_p=0.2t$.
Following Ref. \onlinecite{Monthoux},
the interaction strengths $g$ and $V_{p0}$ are adjusted such that 
the transition temperature becomes $T_{c0}=90$K and the isotope 
exponent reaches $\alpha_0=0.05$. Results for the coupling
constants are shown in Table \ref{Tab1}. Here, the electron-phonon
coupling constant $\lambda_{ph}$ is defined in the usual way:
\begin{equation}
  \lambda_{ph} = 2 \int\limits_0^\infty \frac{d\Omega}{\Omega}
  \; \frac{1}{N} \sum_{q} \frac{1}{\pi} {\rm Im} \; 
  \left\{ V_p(\vec{q},\Omega + i \delta) \right\}
\label{eq19b}
\end{equation}
From Table \ref{Tab1} we can see that higher coupling constants
$g$ are required for smaller spinfluctuation frequencies.
Comparatively small values of the electron-phonon coupling constant 
are sufficient to yield an isotope exponent $\alpha_0=0.05$.

In order to study the influence of a pseudogap we introduce
a $d$-wave pseudogap $ E_g(\vec{k})= E_{g0} \cos 2 \Theta_k$, as 
suggested by the
analysis of Williams et al. \cite{Williams}, into the
single particle excitation spectrum by replacing in the single
particle Green's function Eq. (\ref{eq17})
\begin{equation}
  \epsilon_k + {\rm Re} \Sigma \quad \Longrightarrow \quad 
  \pm \sqrt{(\epsilon_k + {\rm Re} \Sigma )^2 + E_g^2(\vec{k})} .
\label{eq20}
\end{equation}

\begin{table}[b]
\begin{tabular}{cccc}
$\omega_s/t$ & $g/t$ & $\lambda_{ph}$ & $E_{g,{\rm supp}}/T_{c0}$ \\
\hline
0.03 & 5.1 & 0.31 & 12.3 \\
0.06 & 3.2 & 0.15 & 6.5 \\
0.2  & 2.4 & 0.10 & 4.0 
\end{tabular}
\caption{Coupling constants for the NAFL model with an additional
coupling to the buckling mode Eq. (\ref{eq19}) for different values
of the spinfluctuation frequency $\omega_s$. $E_{g,{\rm supp}}$
denotes the value of the pseudogap $E_{g0}$, which completely
suppresses $T_c$.
\label{Tab1} }
\end{table}

Here it is necessary to take into account the real part of the
self-energy $\Sigma$, since the pseudogap opens at the
Fermi surface, which is renormalized due to the self-energy.
Table \ref{Tab1} also shows the amplitude of the pseudogap,
denoted by $E_{g,{\rm supp}}$, which completely suppresses
$T_c$ to 0. Experimentally, the ratio of $E_{g,{\rm supp}}$
to $T_{c0}$ is about 6-15, depending on the material.
\cite{Timusk,Williams} The values found here indeed turn out to
be of this order of magnitude. Note, that the renormalization
of the pseudogap $E_g(\vec{k})$ due to the self-energy $\Sigma$
is taken into account in this approximation. In contrast to
the weak-coupling approximation in the previous section, the
pseudogap as seen in the density of states is washed out now 
and thus is a real 'pseudo'-gap. \cite{DMT}

\begin{figure}[htb]
  \begin{center}
    \includegraphics[width=0.65\columnwidth,angle=270]{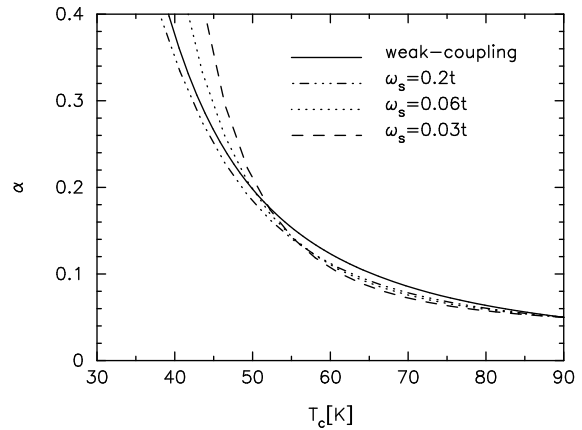}
    \vspace{.2cm}
    \caption{Isotope coefficient $\alpha$ as a function of $T_c$ for the
     NAFL model with an additional coupling to the buckling phonon mode.
     The opening of the pseudogap leads to a suppression of $T_c$ and an
     increase of $\alpha$. The solid line shows the corresponding
     weak-coupling result from Fig. \ref{Fig1}. Results are shown for
     different values of the characteristic spinfluctuation frequency:
     $\omega_s=0.03t$ (dashed line), $\omega_s=0.06t$ (dotted line), 
     and $\omega_s=0.2t$ (dashed-dotted line). In each case the coupling
     constants have been adjusted such that $T_{c0}=90$K and 
     $\alpha_0=0.05$.
     \label{Fig2} }
  \end{center}
\end{figure}

In Fig. \ref{Fig2} we show $\alpha$ as a function of $T_c$ for three
different values of the spinfluctuation frequency $\omega_s$ along
with the weak-coupling result for a $d_{x^2-y^2}$-wave superconductor
with a $d_{x^2-y^2}$-wave pseudogap from Fig. \ref{Fig1}. For higher
values of the spinfluctuation frequency $\alpha$ gradually approaches
the weak-coupling result as one should expect. For small
$\omega_s = 0.03 t$ there are some deviations from the weak-coupling
limit. However, the results are not affected very much.

\section{Self-consistent FLEX approximation}
\label{secflex}

Within the NAFL model the spinfluctuation pairing interaction is fixed
and does not change with the electronic properties. However, it is
clear both experimentally and theoretically that the pseudogap does affect
the spinsusceptibility and thus should affect the spinfluctuation
pairing interaction itself. To study such kind of effects it is
necessary to treat the electronic properties and the electronic
pairing interaction in a self-consistent way. Such a self-consistent
treatment is provided by the so-called fluctuation-exchange (FLEX)
approximation \cite{Bickers} for the two dimensional Hubbard model
and also yields a $d_{x^2-y^2}$-wave superconducting state.
\cite{Pao2,MonScal,Dahm}
The main difference with the NAFL model is that the spinsusceptibility
is calculated from the interacting Green's functions
within an RPA-type approximation. Then the pairing interaction reads:
\begin{equation}
  V(\vec{q},i\nu_m) = \frac{3}{2} U^2 \frac{ \chi_0(\vec{q},i\nu_m)}
{1-U \chi_0(\vec{q},i\nu_m)} \quad ,
\label{eq21}
\end{equation}
where the bubble susceptibility $\chi_0$ is calculated from the fully
dressed single-particle
Green's function $G$ (Eq. (\ref{eq17})) self-consistently:
\begin{equation}
\chi_0(\vec{q},i\nu_m) = -\frac{1}{N} \sum_{k,n}
G(\vec{k}+\vec{q},i\omega_n+i\nu_m)
G(\vec{k},i\omega_n)
\label{eq22}
\end{equation}

\begin{figure}[htb]
  \begin{center}
    \includegraphics[width=0.65\columnwidth,angle=270]{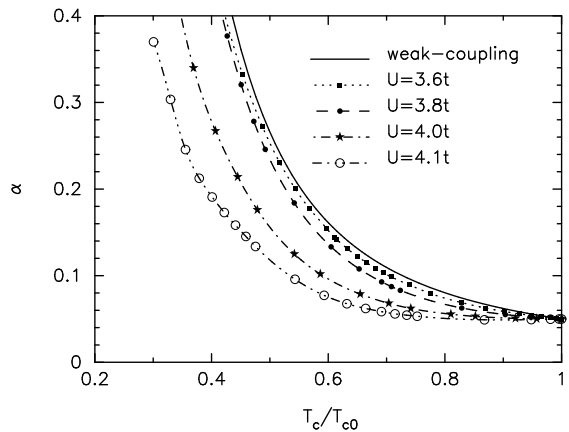}
    \vspace{.2cm}
    \caption{Isotope coefficient $\alpha$ as a function of $T_c/T_{c0}$ within
     FLEX approximation with an additional coupling to the buckling phonon mode.
     Results for
     different values of the on-site Hubbard repulsion $U$ are shown:
     $U=3.6t$ (dotted line), $U=3.8t$ (dashed line), $U=4.0t$ (dashed-dotted line), 
     and $U=4.1t$ (dashed-dot-dotted line). The solid line shows the
     corresponding weak-coupling result from Fig. \ref{Fig1}. 
     \label{Fig3} }
  \end{center}
\end{figure} 

This guarantees that any change in the one particle Green's function is
reflected in the pairing interaction and vice versa. In Fig. \ref{Fig3}
we show $\alpha(T_c/T_{c0})$ within FLEX approximation with an additional
coupling to the buckling mode already considered in the NAFL model above.
Results are shown for different values of the on-site Hubbard
repulsion $U$ along with the weak-coupling result. Here, a bandfilling
of $n=0.84$ for a simple tight-binding band has been assumed. The 
electron-phonon coupling strength again has been adjusted to
give $\alpha_0=0.05$ for $E_{g0}=0$. For $U \le 3.6 t$, $\alpha(T_c/T_{c0})$
very much follows the weak-coupling limit. Only if $U$ reaches values of the
order of $4t$ or higher, deviations from the weak-coupling limit become
apparent. For higher values of $U$, $\alpha(T_c/T_{c0})$ becomes
flatter and starts to rise only at smaller values of $T_c$. This is a
consequence of the influence of the pseudogap on the spinfluctuation
pairing interaction. In contrast to the NAFL model the opening of the
pseudogap not only reduces $T_c$ via the reduction of single
particle spectral weight at the Fermi level, but it also suppresses
the lowest frequency spinfluctuations. Since these are predominantly
pair-breaking \cite{Monthoux2,Millis}, this increases the effective
coupling strength of the spinfluctuations \cite{DahmSSC} as compared 
with the phonons and thus leads to a reduction of $\alpha$.

\section{Conclusions}
\label{seccon}

We studied the influence of a pseudogap on the isotope exponent
for different models having an electronic pairing interaction
with a subdominant electron-phonon interaction. In the
weak-coupling limit we found that the introduction of a
pseudogap leads to a strong increase of the isotope exponent above 
its value in the absence of a pseudogap. For $T_c \rightarrow 0$
the isotope exponent diverges, allowing arbitrarily high
values. The symmetries of the order parameter and the
pseudogap only lead to quantitative, but not qualitative changes
of these results. Strong-coupling effects within the NAFL model
do not affect the results very much. The size of the pseudogap
compared with $T_c$ turns out to be of the right order of
magnitude. Self-consistent treatment
of the spinfluctuation pairing interaction in the presence
of the pseudogap can lead to stronger deviations from the
weak-coupling limit. The general tendency that the isotope
exponent rises upon opening of the pseudogap still remains,
however. From these results it does not seem unreasonable
that the pseudogap indeed can have an important influence on
the isotope effect and might be, at least partially,
responsible for the increasing isotope exponent in the 
underdoped cuprates.

\acknowledgments

The author would like to thank A. Bill, H. Castella, and 
N. Schopohl for valuable discussions and providing 
encouragement for the present study.

\end{document}